\documentclass{vienna-conf2019}
\usepackage{graphicx}
\usepackage{hyperref}
\usepackage[]{natbib}
\usepackage{epstopdf}

\def\BibTeX{{\rm B\kern-.05em{\sc i\kern-.025em b}\kern-.08em
    T\kern-.1667em\lower.7ex\hbox{E}\kern-.125emX}}
\bibpunct{(}{)}{;}{a}{}{,}  
\newcommand{\kms}{{\mathrm{km~s^{-1}}}}

\begin{document}

\TitreGlobal{Stars and their variability observed from space}

\title{Complex asteroseismology of SX Phoenicis}

\runningtitle{Complex asteroseismology of SX Phoenicis}

\author{J. Daszy\'nska-Daszkiewicz}\address{Astronomical Institute,
Wroc{\l}aw University, ul. Kopernika 11, 51--622 Wroc{\l}aw, Poland}

\author{P. Walczak$^1$}

\author{A. Pamyatnykh}\address{Nicolaus Copernicus Astronomical Center,
Polish Academy of Sciences, Bartycka 18, 00--716 Warsaw, Poland}

\author{W. Szewczuk$^1$}

\author{W. Dziembowski$^2$}

\setcounter{page}{237}

\maketitle

\begin{abstract}
We present seismic analysis of the prototype SX Phoenicis that aims at fitting the two radial-mode frequencies and the corresponding values
of the bolometric flux amplitude (the parameter $f$), whose empirical values are derived from multi-coulor photometric observations.
Seismic model that meets these conditions is of low mass, $M=1.05 M_\odot$, has moderately effective convection in the outer layers,
described by the mixing length parameter $\alpha_{\rm MLT} \approx  0.7$,  and the microturbulent  velocity in the atmospheres of about $\xi_{\rm t}\approx 8~\kms$.
Such seismic studies of stars like SX Phe are very important for deriving constraints on outer-layer convection,
because the object is on the border between very effective and ineffective convection.
\end{abstract}


\section{Introduction}

SX Phoenicis  (HD\,223065) is A3-type star of Population II discovered to be variable almost seven decades
ago by \citet{Eggen1952a, Eggen1952b}.
This is a prototype for the whole class of high-amplitude and usually metal-poor pulsators located
inside the $\delta$ Scuti instability region.
SX Phe was a target of several studies based on photometric and spectroscopic observations.
The analysis of photometric data revealed two frequencies and their combinations
\citep[e.g., by][]{Coates1979, Rolland1991, Garrido1996}.
The frequency ratio indicated that SX Phe pulsates in the two radial modes; fundamental and first overtone ones.
These two periodicities are present also in the radial velocity variations \citep{Kim1993}.
Interestingly, the recent analysis of the high-precision data from the TESS satellite confirmed the old results
that only these two frequencies
dominate the light curve of SX Phe \citep{Antoci2019}. The values of frequencies extracted from the TESS light curve
are $\nu_1=18.193565(6) ~{\mathrm{d^{-1}}}$ and $\nu_2=23.37928(2) ~{\mathrm{d^{-1}}}$.

There is also some evidence that both pulsation periods change in a timescale of decades \citep{Landes2007}.
Moreover, for the dominant pulsational period the effective temperature varies in a huge range
from 7210 to 8120\,K and the surface
gravity from 3.63 to 4.23. The corresponding mean values are 7640 K and 3.89 \citep{Kim1993}.
The recent determination of the effective temperature from spectroscopy amounts
to  $T_{\rm eff} = 7500\pm 150$\,K
and the luminosity derived from the Gaia DR2 data is  $\log L/L_\odot=0.844\pm0.009$ \citep{Antoci2019}.
The metallicity of SX Phe is low and amounts to about [Fe/H]$=-1.4$ \citep{McNamara1997}.

SX Phe has been also a subject of seismic modelling. In the first attempts,
\citet{Dziembowski1974} indicated a small mass of about 0.2\,$M_\odot$.
The modeling with the recomputed OPAL opacity data \citep{Iglesias1992}
by \citet{Petersen1996}
showed that the period ratio of the two radial modes is best reproduced by the model with parameters:
a mass $M= 1.0M_\odot$, metallicity $Z = 0.001$, initial hydrogen abundance $X_0 = 0.70$ and age 4.07\,Gyr.

The aim of this paper is to repeat seismic modelling of SX Phe and extend it by employing
the bolometric flux amplitude (the parameter $f$)
which is very sensitive to physical conditions in subphotospheric layers.

\section{Models fitting the two radial-mode frequencies}

Using the Warsaw-New Jersey evolutionary code \citep[e.g.,][]{Pamyatnykh1998, Pamyatnykh1999}
and nonadiabatic pulsational code \citep{Dziembowski1977}, we computed models of SX Phe with
the aim to fit $\nu_1$ and $\nu_2$ as the radial fundamental and first overtone modes, respectively.
Part of these results have been already published in \citet{Antoci2019}.

We searched the whole range of the effective temperature derived for SX Phe in the literature, i.e.,  from about 7000 to 9000\,K.
The OPAL opacity tables and the solar element mixture of \citet{Asplund2009} were adopted.
We considered different values of the metallicity, $Z$, and initial hydrogen abundance, $X_0$,
from the range (0.0005, 0.003) and (0.66, 0.74), respectively.
Different values of the mixing length parameter were investigated $\alpha_{\rm MLT} \in (0,~2.5)$
and overshooting from the convective core was not included.
At such low metallicity and effective temperatures $T_{\rm eff}>7000$\,K, the effect of $\alpha_{\rm MLT}$
on mode frequencies is negligible
but it strongly affects the values of the parameter $f$, discussed in the next section.


We started from re-calculation of the seismic model of \citet{Petersen1996},
whose constraints were: a mass $M = 1.0~M_\odot$,  metallicity $Z = 0.001$ and initial hydrogen $X_0 = 0.70$.
Our best seismic model for these parameters has the effective temperature $\log T_{\rm eff}=3.88427$,
luminosity $\log L/L_{\odot}=0.811$
and the frequency ratio of the radial fundamental mode to the first overtone amounts to $\nu_1/\nu_2=0.780014$.
The observed counterpart is 0.778192.
The theoretical values of the frequencies are $\nu_1=18.193565$ d$^{-1}$ (the model was interpolated
on this frequency)
and  $\nu_2=23.324654$ d$^{-1}$.  Both radial modes are in this model excited. However, as one can see,
the difference between the theoretical and observed value of $\nu_1/\nu_2$ is significant and amounts to 0.001822.
The age of the model is about 3.93 Gyr, which is slightly less than for the model of \citet{Petersen1996} who obtained about 4.07 Gyr. 
This difference may results from different versions of the OPAL opacity tables as well we from the usage of various evolutionary
and pulsational codes.

In the next step,  we changed the value of the metallicity and found the model for $Z=0.0014$
with a mass $M=1.15~M_\odot$ which fits better the observed frequency ratio.
The model has the following parameters: $\log T_{\rm eff}=3.91770$, luminosity $\log L/L_{\odot}=0.984$.
The frequency ratio is $\nu_1/\nu_2=0.778412$ and the individual frequencies are $\nu_1=18.193565$ d$^{-1}$
and  $\nu_2=23.372681$ d$^{-1}$. The model is much younger than the previous one with the age of about 2.45 Gyr.
The fundamental and first overtone radial modes are in this model both stable. 
Increasing slightly metallicity, to $Z=0.002$ allowed to find the model with a mass $M=1.05~M_\odot$,
$\log T_{\rm eff}=3.87152$, $\log L/L_{\odot}=0.770$ and with the frequency ratio of the two considered radial modes  $\nu_1/\nu_2=0.778073$.
This models is about 3.42 Gyr old and have both radial fundamental and first overtone modes excited.
As one can see the models with higher metallicity abundance give much better fit of frequencies but still there is a room for an improving.
Therefore, in the next step we changed also the initial abundance of hydrogen.

We have managed to find the model that reproduces the observed frequency ratio up to the fifth decimal place
which we adopt as the numerical accuracy.
The model has the following parameters: $M=1.05~M_\odot$, $X_0=0.67$, $Z=0.002$, $\log T_{\rm eff}=3.88979$,
$\log L/L_{\odot}=0.844$
and it was interpolated on the dominant frequency $\nu_1=18.193565$ d$^{-1}$.  The value of the second frequency
is  $\nu_2=23.379761$ d$^{-1}$  which differs by  0.00048 d$^{-1}$ from the observed value.
Thus, the theoretical value of the frequency ratio is  0.778176  compared to the observed value 0.778192.
Taking into account the numerical accuracy, which is not better than five decimal places, we can conclude
that this model perfectly reproduces the observed frequencies of the two radial modes of SX Phe.
Moreover, this model predicts instability (excitation) of both radial fundamental and first overtone modes.
The age of the model amounts to about 2.85 Gyr.
Another advantage of this model 
is the fact that it has the luminosity which agrees with the value derived from the Gaia DR2 data.

The lower hydrogen abundance (higher helium abundance) is also quite probable because, as
other SX Phoenicis variables,
SX Phe itself can be a blue straggler. Such objects are presumably formed by the merger of two stars
or by interactions in a binary system and, as a consequence, they may have
enhanced helium abundance \citep[e.g.,][]{McNamara2011, Nemec2017}.
\begin{figure}[ht!]
 \centering
 \includegraphics[width=0.49\textwidth,clip]{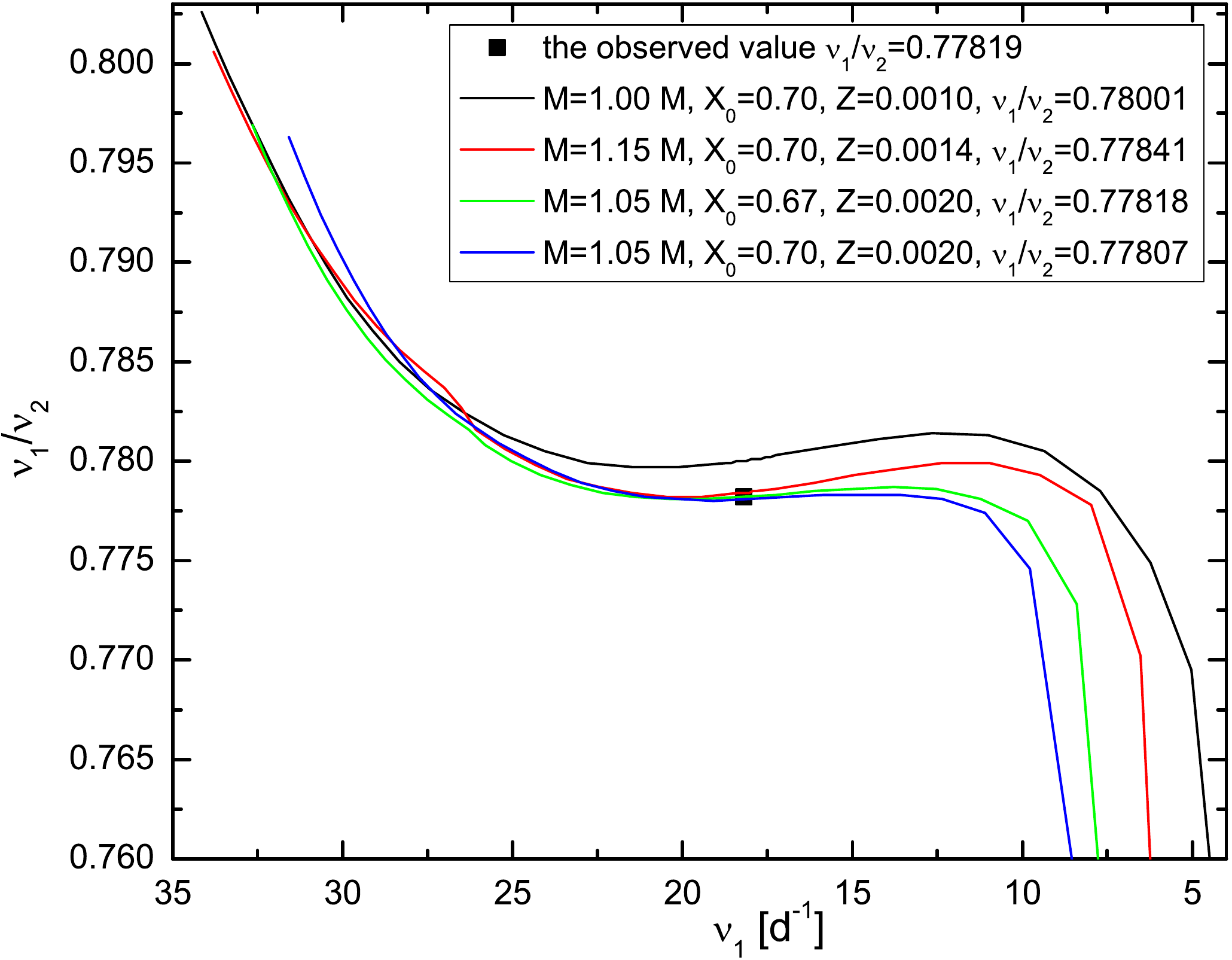}
 \includegraphics[width=0.48\textwidth,clip]{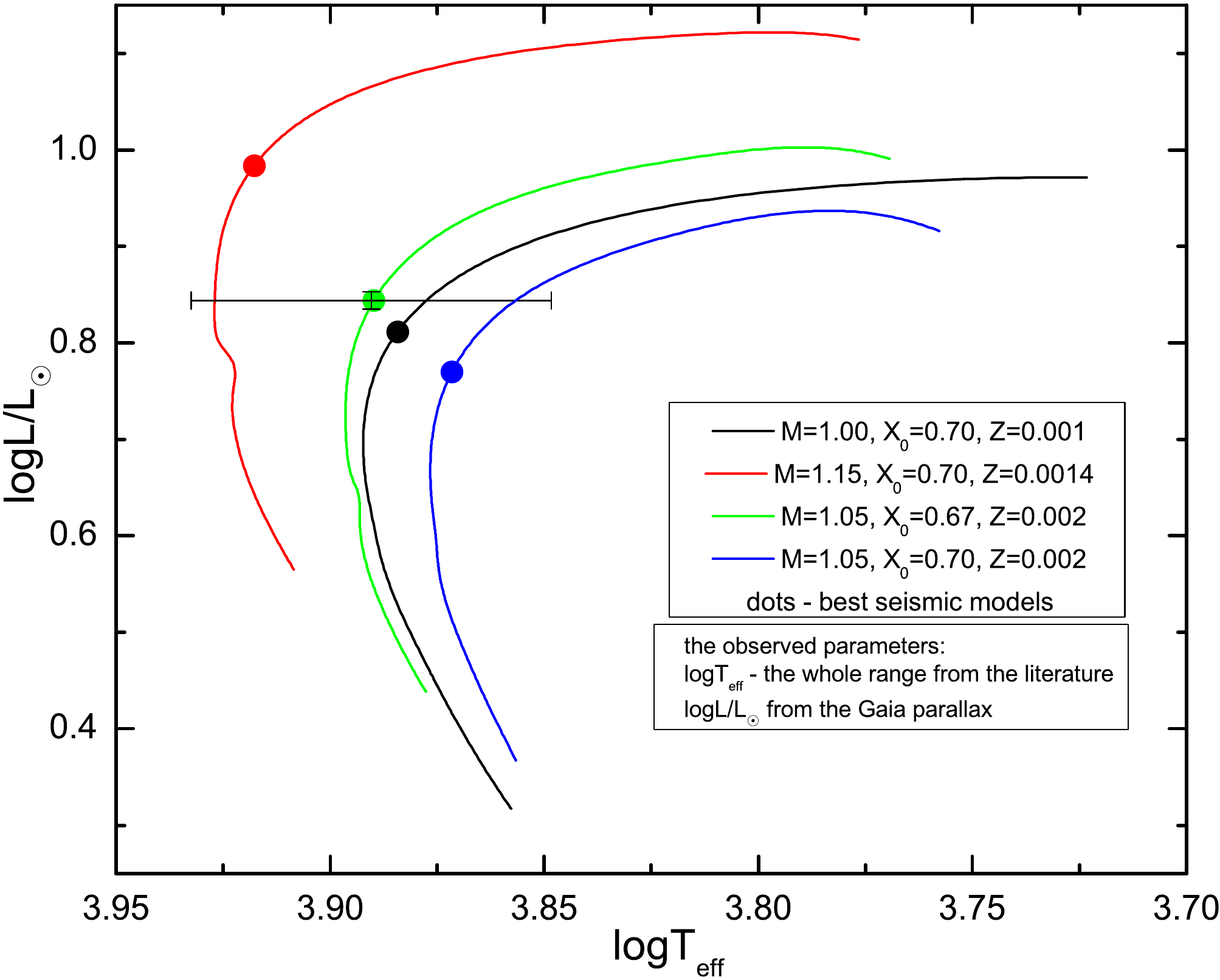}
  \caption{{\bf Left:} The Petersen diagram for the four best seismic models suitable for SX Phoenicis. The solid square marks the observed values.
  The theoretical values of $\nu_1/\nu_2$ are given in the legend along with the mass and chemical composition $(X,Z)$.
   {\bf Right:} The corresponding evolutionary tracks of the seismic models with the observed position of SX Phe. }
  \label{fig1}
\end{figure}

In the left panel of Fig.\,1, we show the evolution of the frequency ratio as a function of the dominant
frequency (the Petersen diagram) for the four best seismic models we have found. The observed value is marked as a solid square.
In the right panel of Fig.\,1, we depicted the corresponding evolutionary tracks on the HR diagram and the position of SX Phe.

\section{Complex seismic modeling}

In the next step we extended our seismic analysis by including the parameter which describes the relative amplitude of the radiative flux perturbations at the photosphere level.
This amplitude is the so-called nonadiabatic parameter $f$ and it is defined as
\begin{equation}
\frac{ \delta {\cal F}_{\rm bol} } { {\cal F}_{\rm bol} }= {\rm Re}\{ \varepsilon f Y_\ell^m(\theta,\varphi) {\rm e}^{-{\rm i} \omega t} \},
\label{eq3}
\end{equation}
where ${\cal F}_{\rm bol}$ is the bolometric flux, $\varepsilon$ is the mode intrinsic amplitude, $Y_\ell^m$ is the spherical harmonic with
the degree $\ell$ and the azimuthal order $m$.  Other quantities have their usual meaning. The value of $f$ is complex and it is associated with a given mode.
The theoretical values of $f$ are very sensitive to the subphotospheric layers where the driving of pulsation occurs.
In the case of B-type pulsators, these values depend mainly on the adopted opacity data and their modifications around the Z-bump \citep{JDD2005}.
In case of cooler pulsators, like $\delta$ Scuti or SX Phoenicis stars, the values of $f$ are affected by convection in the envelope \citep{JDD2003}.

Therefore, in this paper we aim at obtaining some constraints on the efficiency of convective transport in the outer layers of SX Phe.
The semi-empirical values of $f$ can be derived from multi-colour light variations and radial velocity measurements \citep[e.g.,][]{JDD2003, JDD2005}.
To this end, we used the Str\"omgren amplitudes and phases for the two radial mode frequencies as determined by \citet{Rolland1991}.
The atmospheric flux derivatives over the effective temperature and gravity as well as limb darkening were derived 
from the Vienna model atmospheres  \citep[NEMO, see, e.g.,][]{Nendwich2004}.

In Fig.\,2 we show a comparison of the theoretical and empirical values of $f$ for our best seismic model 
with the parameters: $M=1.05~M_{\odot}$, $\log T_{\rm eff}=3.889793$, $\log L/L_{\odot}=0.84375$ and $X_0=0.67$ and $Z=0.002$.
The theoretical values were computed for different values of the mixing length parameter $\alpha_{\rm MLT}$  
whereas the empirical counterparts for different values of the microturbulent velocity $\xi_t$. In the left panel, 
we plotted the results for the dominant frequency $\nu_1$ (the radial fundamental mode) 
and in the right panel for the frequency $\nu_2$ (the first overtone radial mode).
\begin{figure}[ht!]
 \centering
 \includegraphics[width=0.48\textwidth,clip]{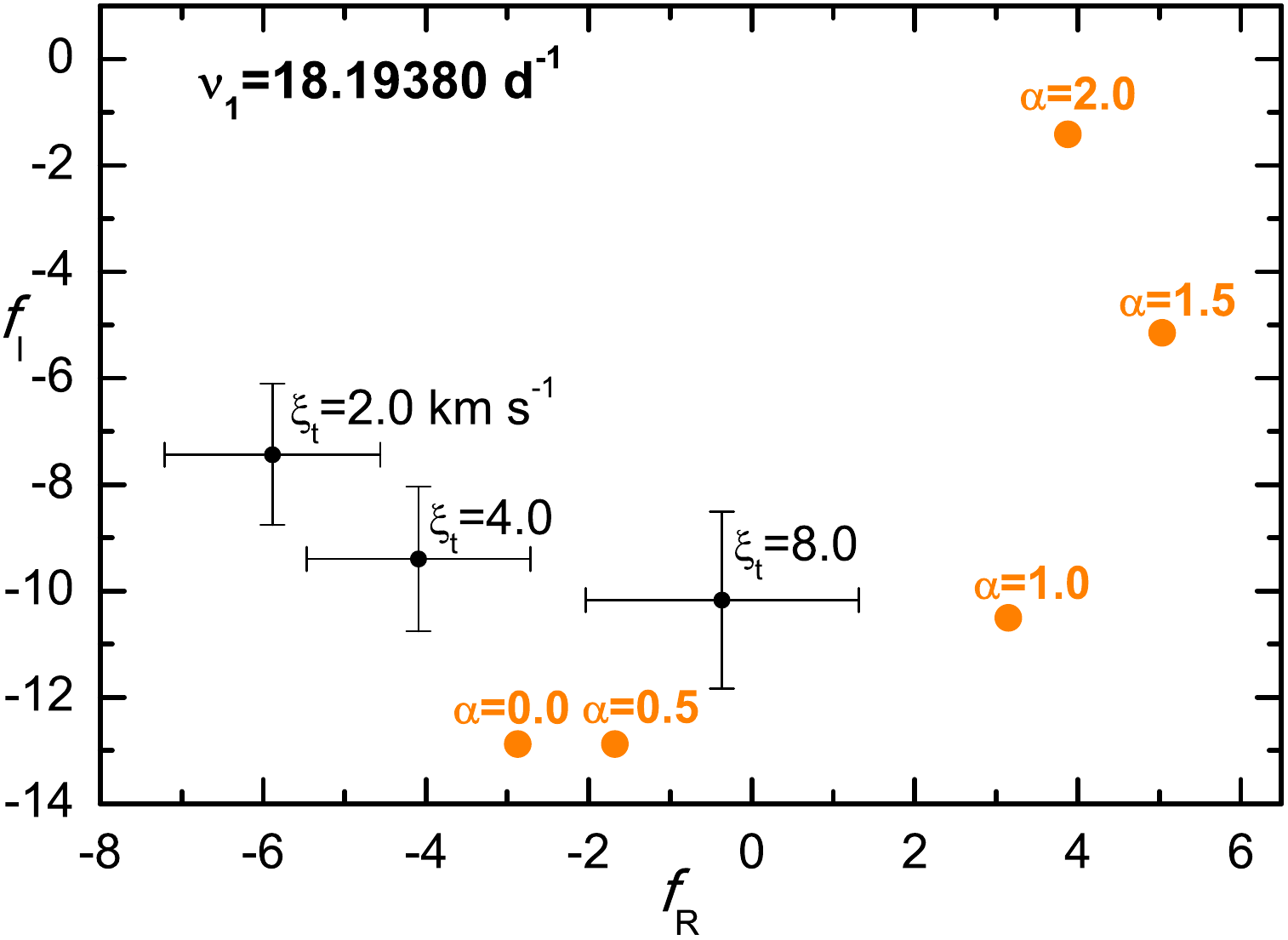}
 \includegraphics[width=0.49\textwidth,clip]{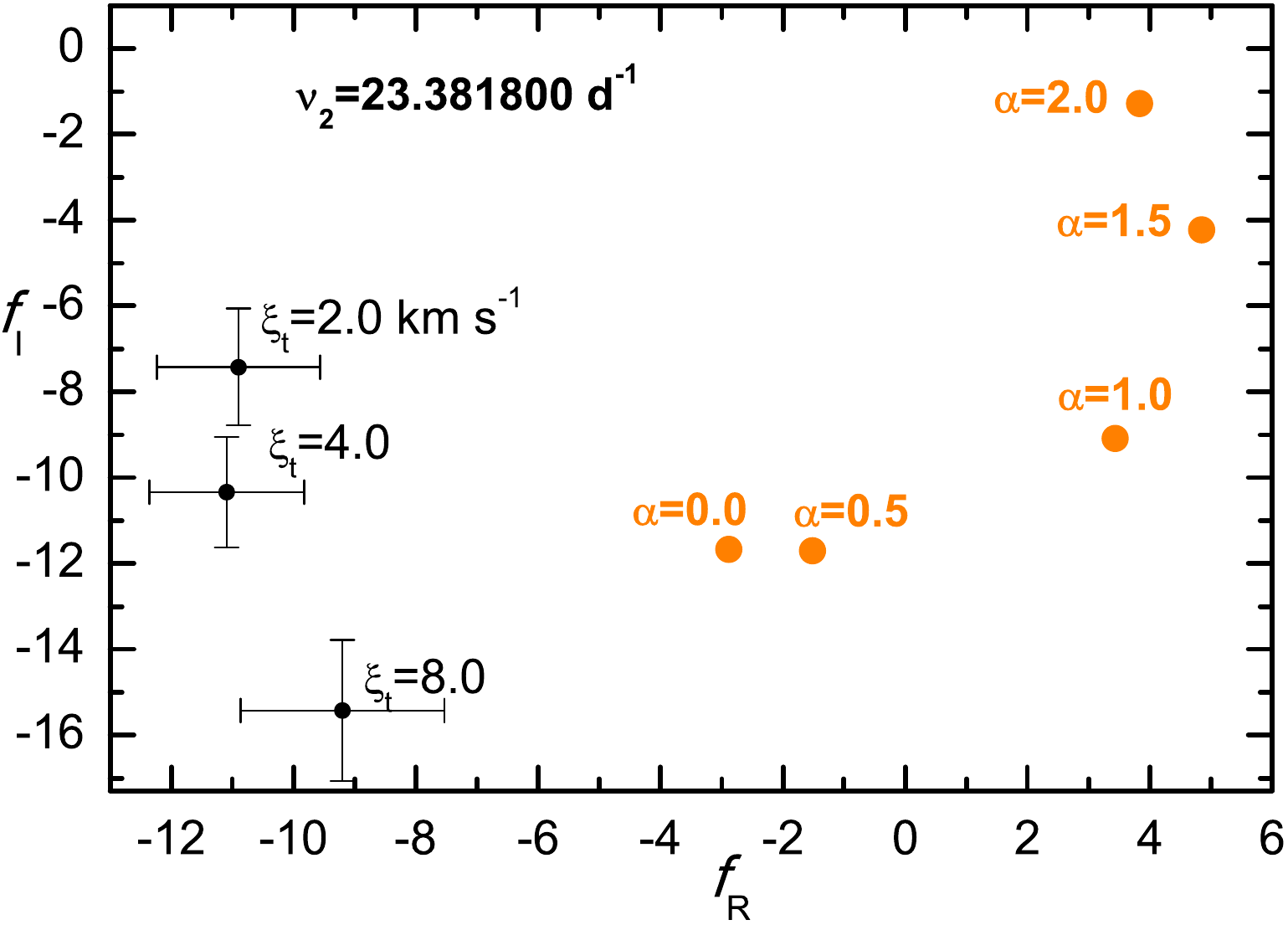}
  \caption{A comparison of the theoretical and empirical values of $f$ on the complex plane for the radial fundamental mode (the left panel)
   and for the first overtone mode (the right panel). The theoretical values were computed for our best seismic model (see text for details)
   considering different values of the mixing length parameter $\alpha_{\rm MLT}$. The (semi)empirical counterparts were determined from the Str\"omgren photometry
   and the Vienna model atmospheres assuming different values of the microturbulent velocity $\xi_t$.}
  \label{fig2}
\end{figure}

As one can see, in the case of the frequency $\nu_1$ there is a pretty good agreement for the MLT parameter of $\alpha_{\rm MLT}\approx 0.7$ 
and the microturbulent velocity of $\xi_{\rm t}\approx 8~\kms$.  The results for the second mode frequency $\nu_2$ are less unambiguous.
While the imaginary part of $f$ indicates lower values of the MLT parameter ($\alpha_{\rm MLT}<1.5$), the real part of $f$ does not agree with any theoretical value.
It can results from much smaller light amplitude of $\nu_2$ which is determined with much lower accuracy. For example, the amplitude in the Str\"omgren $v$ filter
is almost three times lower for the frequency $\nu_2$ than for the frequency $\nu_1$.

\section{Conclusions}

We made seismic analysis of the prototype SX Phoenicis. We started from the fitting of the two radial mode frequencies and constrained the mass, 
luminosity and chemical composition.
Our best seismic model, that reproduces these two frequencies has the parameters: $M=1.05~M_{\odot}$, $\log T_{\rm eff}=3.889793$, $\log L/L_{\odot}=0.84375$
and the chemical composition: $X_0=0.67$ and $Z=0.002$. The effective temperature is within the allowed observational range and luminosity perfectly agrees 
with the determination from the Gaia DR2 data. 
The age of other seismic models found by us is in the range from 2.5 to 3.9 Gyr.
Further studies are needed to estimate the age more accurately, because it would give a clue to the star's evolutionary past and its origin. 

In the next step, we tried to reproduce also the bolometric flux amplitude (the parameter $f$) corresponding to each mode. The aim was to get further constraints, 
e.g., on convection or atmospheric conditions.
The (semi)empirical values of $f$ were derived from the Str\"omgren amplitudes and phases adopting the Vienna model atmospheres.
We found a very strong effect of the microturbulent velocity, $\xi_t$, on the empirical values of $f$ which was already announced by \citet{JDD2005a}
and \citet{JDD2007}.
In the case of the fundamental radial mode, the empirical and theoretical values of $f$ agree if the MLT parameter is about $\alpha_{\rm MLT}=  0.7$
and the microturbulent velocities in the atmospheres is about $\xi_t=  8~\kms$.  It would mean that the efficiency of convective transport in the outer layers of SX Phe is rather moderate.
In the case of the first overtone mode the agreement  is poor and further studies are needed. It can result from some interaction between the two modes
or a need of additional modification in pulsational and/or atmospheric models. Another reason could be much smaller photometric amplitudes determined with much lower accuracy. 

We are going to extend these studies by re-computations of atmospheric models for a higher abundance of helium and higher microturbulent velocities.
Moreover, the effect of modification of the mean opacities will be examined.

Definitely, new simultaneous multi-colour photometric and spectroscopic time series observations would lead to more plausible seismic constraints
on the parameters of the model and theory.


\begin{acknowledgements}
The work was financially supported by the Polish NCN grant 2018/29/B/ST9/02803.
\end{acknowledgements}

\bibliographystyle{aa}  
\bibliography{JDaszynskaD_3o02} 

\end{document}